\documentclass{article}

\usepackage{amsmath}
\usepackage{amssymb}
\usepackage{dsfont}

\newcommand{\Xdd}{\ddot{X}}
\newcommand{\Xd}{\dot{X}}
\renewcommand{\Xi}{X^{-1}}
\newcommand{\Ri}{R^{-1}}
\newcommand{\Rd}{\dot{R}}
\newcommand{\Sd}{\dot{s}}
\newcommand{\Sdd}{\ddot{s}}
\newcommand{\Ld}{\dot{L}}
\newcommand{\Dd}{\dot{D}}
\newcommand{\Gd}{\dot{G}}
\newcommand{\Qd}{\dot{Q}}
\newcommand{\Qv}{\vec{\,Q}}

\newcommand{\Pd}{\dot{P}}
\newcommand{\Pv}{\vec{P}}

\newcommand{\qd}{\dot{q}}
\newcommand{\qdd}{\ddot{q}}
\newcommand{\qv}{\vec{q}}
\newcommand{\pd}{\dot{p}}
\newcommand{\pv}{\vec{p}}
\newcommand{\fd}{\dot{f}}
\newcommand{\uv}{\vec{u}}
\newcommand{\uvd}{\dot{\uv}}
\newcommand{\vv}{\vec{v}}
\newcommand{\vvd}{\dot{\vv}}
\newcommand{\xv}{\vec{x}}
\newcommand{\xd}{\dot{x}}
\newcommand{\xdd}{\ddot{x}}
\newcommand{\piv}{\vec{\pi}}
\newcommand{\pit}{\tilde{\pi}}
\newcommand{\pid}{\dot{\pi}}
\newcommand{\ld}{\dot{\lambda}}
\newcommand{\ldd}{\ddot{\lambda}}

\newcommand{\Tr}{\operatorname{Tr}}
\newcommand{\half}{\frac{1}{2}}
\renewcommand{\d}{\partial}
\renewcommand{\mid}{\mathds{1}}

\newcommand{\para}[1]{\left(#1\right)}
\newcommand{\pb}[1]{\left\{#1\right\}}

\newcommand{\sumt}{\overset{\sim}{\sum}}
\newcommand{\sump}[1]{{\sum_{#1}}^{'}}

\title{\raggedright \textbf{Goldfish geodesics and \\ Hamiltonian reduction of matrix dynamics}}
\author{\raggedright Joakim Arnlind$^{\ast}$, Martin Bordemann, Jens Hoppe, Choonkyn Lee}
\date{}

\begin{document}

{\raggedright\LARGE\textbf{Goldfish geodesics and Hamiltonian \\ reduction of matrix dynamics}
\vspace{0.7cm}\\
{\normalsize Joakim Arnlind$^{\ast}$, Martin Bordemann$^\dagger$, \\Jens Hoppe$^{\ast\sim}$, Choonkyu Lee$^\sim$}
}
\vspace{0.5cm}\\
{\scriptsize
\noindent $^{\ast}$ Department of Mathematics, Royal Institute of Technology, 100 44 Stockholm, Sweden.\\
$^\dagger$ Laboratoire de MIA, 4, rue des Fr{\`e}res Lumi{\`e}re, Univ. de Haute-Alsace, F-68093 Mulhouse.\\
$^\sim$ Dept. of Physics and Center for Theoretical physics, Seoul National Univ., Seoul 151-742, Korea.
}

\vspace{0.6cm}

\noindent \textbf{Abstract}\vspace{0.1cm}\\
\noindent We relate free vector dynamics to the eigenvalue motion of a
time-dependent real-symmetric $N\times N$ matrix, and give a
geodesic interpretation to Ruijsenaars Schneider models.

\vspace{0.8cm}

\noindent Despite of more than two decades of extensive work on
Ruijsenaars-Schneider models \cite{Calogero:1978,RS:1986}, their
interpretation as describing geodesic motions seems to have gone
unnoticed. Apart from wishing to fill this gap, the (related) second
topic of this paper is the Hamiltonian reduction of free matrix
dynamics (to, as we will show, free vector dynamics). 

To start with the latter, let $\Xdd(t)=0$, $X(t)$ a real symmetric
$N\times N$ matrix (depending on ''time''), be described by
\begin{align}\label{eq:freeHamiltonian}
  H[X,P]:=\half\Tr P^2.
\end{align}
Writing
\begin{align}
  \begin{split}
    X(t) &= R(t)D(t)\Ri(t)\\
    \Xd(t) &= R\para{\Dd+[M,D]}\Ri=RL\Ri,
  \end{split}
\end{align}
the symplectic form $-\Tr dX\wedge dP$ becomes (cp. \cite{AH:LesHouches}), with $dA:=\Ri dR$,
\begin{equation}\label{eq:sympqpfa}
  \begin{split}
    -dq_i\wedge &dp_i+2\sum_{i<j}df_{ij}\wedge da_{ij}\\
    &-2\sum_{i<j<k}\para{f_{ij}da_{jk}\wedge da_{ik}+f_{ik}da_{ij}\wedge da_{jk}+f_{jk}da_{ik}\wedge da_{ij}},
  \end{split}
\end{equation}
which (inverting \eqref{eq:sympqpfa}) gives the non-trivial Poisson-brackets
\begin{align}
  \pb{q_i,p_j} &= \delta_{ij}\label{eq:pbqp}\\
  \pb{f_{ij},f_{kl}}&=-\half\delta_{jk}f_{il}+\half\delta_{ik}f_{jl}+\half\delta_{jl}f_{ik}-\half\delta_{il}f_{jk}\label{eq:pbff}\\
  %\pb{f_{i<j},f_{j<k}} &= -\half f_{i<k}
  \pb{f_{i<j},a_{k<l}} &= -\half \delta_{ik}\delta_{jl}\label{eq:pbfa}\\
  \text{resp. }\pb{r_{ij},f_{kl}} &= - \half\para{\delta_{jk}r_{il}-\delta_{jl}r_{ik}}\label{eq:pbrf}
\end{align}
for the eigenvalues of $X$, their time derivatives $\qd_i$, and\\
$f_{ij}=-f_{ji}:=\para{\Ri \Rd}_{ij}(q_i-q_j)^2$.
\eqref{eq:freeHamiltonian} becomes
\begin{equation}\label{eq:qpfHamiltonian}
  H = \half\sum_{i=1}^N p_i^2 +\half\sum_{i\neq j}\frac{f_{ij}^2}{(q_i-q_j)^2},
\end{equation}
which, as $H=H[\qv,\pv;f_{ij}]$, is known under the name ''Euler
Calogero-Moser Hamiltonian'' (\cite{Gibbons:1984,Wojciechowski:1985}),
with equations of motion
\begin{equation}\label{eq:qfmotion}
  \begin{split}
    \qdd_i &= 2\sum_{k\neq i} \frac{f_{ik}^2}{(q_i-q_k)^3}\\
    \fd_{ij} &= -\sum_{k\neq i,j}f_{ik}f_{kj}\para{\frac{1}{q_{ik}^2}-\frac{1}{q_{kj}^2}}.
  \end{split}
\end{equation}
While it is well known (see e.g. \cite{Calogero:2,AH:2004}) that both (types of)
equations in \eqref{eq:qfmotion} consistently reduce to
\begin{equation}\label{eq:goldfish}
  \qdd_i = 2\sum_{j(\neq i)}\frac{\qd_i\qd_j}{q_i-q_j}
\end{equation}
upon setting
\begin{equation}
  f_{ij} = -(q_i-q_j)\sqrt{\qd_i\qd_j},
\end{equation}
the \emph{Hamiltonian} reduction of \eqref{eq:freeHamiltonian} (resp.
\eqref{eq:qpfHamiltonian}) to \eqref{eq:goldfish} has remained open
for many years. Simply counting the degrees of freedom
($H[\qv,\pv;f_{i<j}]$ has at most $2N+\frac{N(N-1)}{2}$; to obtain a
nice phase-space one would have to eliminate the degrees of freedom
corresponding to the Casimirs of the SO(N) generated by the $f_{ij}$),
together with
\begin{equation}\label{eq:Gdef}
  \begin{split}
    \pb{G_{ij},G_{kl}}&= -\delta_{jk}G_{il}+\delta_{ik}G_{jl}+\delta_{jl}G_{ik}-\delta_{il}G_{jk}\\
    G_{ij} &:= 2\para{f_{ij}+(q_i-q_j)\sqrt{p_i p_j}}
  \end{split}
\end{equation}
shows that one can not simply ''fix the gauge''. Viewing
\eqref{eq:qpfHamiltonian} as originating from
\eqref{eq:freeHamiltonian}, however, one has the
degrees of freedom corresponding to the orthogonal matrix $R$ (resp.
the antisymmetric matrix $A$, which -- due to \eqref{eq:pbfa} --
naturally provides $N(N-1)/2$ variables that \emph{are} canonically
conjugate to the $G_{ij}$). Note that \eqref{eq:qpfHamiltonian},
when written in terms of the variables $G_{ij}$, reads
\begin{equation}\label{eq:pqGHamiltonian}
  H = \half\para{\sum_{i=1}^N p_i}^2
  +\frac{1}{8}\sum_{i\neq j}\frac{G_{ij}^2}{(q_i-q_j)^2}
  -\frac{1}{4}\sum_{i\neq j}\frac{G_{ij}\sqrt{p_i p_j}}{q_i-q_j},
\end{equation}
clarifying (in the Hamiltonian framework) that it is consistent to put
the $G_{ij}=0$, due to $\Gd_{ij}=\pb{G_{ij},H}$ being ''weakly zero''
(i.e. using $G_{kl}=0$ after computing the Poisson-brackets, according
to \eqref{eq:Gdef} and \eqref{eq:pbqp}/\eqref{eq:pbff}; note that
while the $q$'s and $p$'s do not commute with the $G_{ij}$, the total
momentum, $P:=\sum p_i$, \emph{does}, as the $G_{ij}$ contain only the
differences of the $q$'s). In order to find out what the reduced phase
space is (consisting of functions that Poisson-commute with all the
$G_{ij}$), the reduced Hamiltonian (cp. \eqref{eq:pqGHamiltonian}) being
\begin{equation}\label{eq:HP}
  H = \half\para{\sum p_i}^2 = \half P^2,
\end{equation}
it is useful to bring the ''orbital angular momentum'',
\begin{equation}
  L_{ij} := 2(q_i-q_j)\sqrt{p_i p_j}
\end{equation}
(satisfying \eqref{eq:Gdef}, with $G$ replaced by $L$) into canonical
form ($Q_i P_j-Q_j P_i$), by making the canonical transformation 
\begin{equation}
  P_i = \sqrt{p_i},\qquad Q_i=2q_i\sqrt{p_i}.
\end{equation}
While the (first order form, $\qd_i=p_i$, $\pd_i=\ldots$, of the)
goldfish-equations \eqref{eq:goldfish} do(es) not simplify at all,
\begin{equation}\label{eq:QPmotion}
  \begin{split}
    &\Qd_i = 2P_i^3+2Q_i P_i\sum_j \frac{P_j^3}{Q_i P_j - Q_j P_i}\\
    &\Pd_i = 2P_i^2\sum_j\frac{P_j^3}{Q_i P_j - Q_j P_i}\\
  \end{split}
\end{equation}
(with $\Qd_i P_i - Q_i\Pd_i = 2P_i^4$) $\Qv$ and $\Pv$ now transform as ordinary vectors ($\Qv\to S\Qv$, $\Pv\to
S\Pv$) under the rotations generated by the $G_{ij}$:
\begin{equation}
  \begin{split}
    \pb{G_{ij},Q_k} &= \delta_{ik}Q_j-\delta_{jk}Q_i\\
    \pb{G_{ij},P_k} &= \delta_{ik}P_j-\delta_{jk}P_i.
  \end{split}
\end{equation}
Together with (cp. \eqref{eq:pbrf})
\begin{equation}
  \pb{G_{ij},r_{kl}} = \delta_{il}r_{kj}-\delta_{jl}r_{ki},
\end{equation}
corresponding to $R\to RS$, the $2N$ independent variables 
\begin{equation}
  \uv := R\Qv,\qquad \vv:=R\Pv
\end{equation}
are therefore invariant, i.e. natural coordinates
for the reduced phase-space. The original free matrix-dynamics (with
the initial condition that $\Xd(0)$ has 1 positive eigenvalue, and
$N-1$ zero) is thereby reduced to free vector dynamics,
\begin{equation}
  \uvd = 2\vv\para{\vv^2},\qquad \vvd = 0,
\end{equation}
governed by
\begin{equation}
  H[\uv,\vv] = \half\para{\vv^{\,2}}^2\,\,
  \para{=\half\para{(R\Pv)^2}^2=\half(\Pv^2)^2=\half P^2}.
\end{equation}
Using \eqref{eq:QPmotion}, and the time-evolution of the $r_{ij}$ (cp.
\eqref{eq:pbrf}/\eqref{eq:qpfHamiltonian}) given according to 
\begin{equation}
  \para{\Ri\Rd}_{ij} = M_{ij} = \frac{f_{ij}}{q_{ij}^2}=
  -\frac{\sqrt{p_i p_j}}{q_{ij}}=-2\frac{P_i^2 P_j^2}{Q_i P_j-Q_jP_i}
\end{equation}
one can check the time-evolution of the invariant variables $\uv=R\Qv,\vv=R\Pv$:
\begin{equation}
  \frac{d}{dt}\para{R\Qv} = 2\para{R\Pv}\Pv^2,\quad 
  \frac{d}{dt}\para{R\Pv}=0.
\end{equation}
Let us know focus on \eqref{eq:goldfish}, independent of any previous
considerations, as (coupled) second order ODE's. Their most
natural, and extremely simple (though apparently unnoticed),
interpretation is that of \emph{geodesic} equations,
\begin{align}
  &\qdd^i + \Gamma_{jk}^i\qd^j\qd^k=0\label{eq:geodesic}\\
  &\Gamma_{jk}^i = -\para{\frac{\delta_j^i(1-\delta_k^i)}{q^i-q^k}+\frac{\delta_k^i(1-\delta_j^i)}{q^i-q^j}}.\label{eq:Christoffel}
\end{align}
The simple expression \eqref{eq:Christoffel}, which is of the form
\begin{equation}
  \Gamma_{jk}^i = \delta_j^i w_{ik}+\delta_k^i w_{ij},
\end{equation}
with $w_{ik}=-\half w(q^i-q^k)(1-\delta_{ik})$, makes the calculation of the curvature-tensor
$R_{adb}^c=\d_d\Gamma_{ab}^c+\Gamma_{ab}^e\Gamma_{de}^c-(b\leftrightarrow d)$ straightforward:
\begin{equation}\label{eq:Curvaturetensor}
  \begin{split}
    R_{adb}^c = \,&\delta_a^c\delta_d^c w'_{ab}-\delta_a^c\delta_b^c w'_{ad}+w'_{ca}\para{\delta_{ab}\delta_d^c-\delta_b^c\delta_{da}}\\
    &+\delta_d^c\para{w_{ab}w_{ca}+w_{ba}w_{cb}-w_{ca}w_{cb}}\\
    &-\delta_{b}^c\para{w_{ad}w_{ca}+w_{da}w_{cd}-w_{ca}w_{cd}}
  \end{split}
\end{equation}
is identically zero for $w(x)=2/x$ (i.e. \eqref{eq:Christoffel}), due to
\begin{equation}
  \begin{split}
    \frac{1}{q_c-q_a}\frac{1}{q_a-q_d}+\frac{1}{q_a-q_d}\frac{1}{q_d-q_c}+\frac{1}{q_d-q_c}\frac{1}{q_c-q_a}=0\qquad(a\neq c\neq d \neq a),
  \end{split}
\end{equation}
and the $w'$ term being cancelled by those $ww$ terms, for which an
additional equality of indices holds; alternatively, one could e.g. bring
the last 3 $ww$-terms in \eqref{eq:Curvaturetensor}, provided $w(-x)=-w(x)$, into the form
\begin{equation}
  w(x)w(y)-w(x+y)\para{w(x)+w(y)},
\end{equation}
$x=q_c-q_a,y=q_a-q_d$, and then note the linearity of $1/w$.

In order to find the change of variables that will make the
Christoffel-symbols vanish, it is easiest to observe that
\begin{equation*}
  \qdd_i = 2{\sum_j}'\frac{\qd_i\qd_j}{q_i-q_j}
\end{equation*}
directly implies that the functions
\begin{equation}\label{eq:bn}
  b_n(t):=\frac{1}{(n-1)!}\sumt q_{i_1}q_{i_2}\cdots q_{i_{n-1}}\qd_{i_n}
  \qquad n=1,2,\ldots
\end{equation}
(with $\sim$ indicating the indices to all be different) do \emph{not} depend on time:
\begin{equation}
  \begin{split}
    &\frac{1}{(n-2)!}\sumt q_{i_1}q_{i_2}\cdots \qd_{i_{n-1}}\qd_{i_n}\\
    &+\frac{2}{(n-1)!}\sumt q_{i_1}q_{i_2}\cdots q_{i_{n-1}}
    \sum_{j\neq i_n}\frac{\qd_{i_n}\qd_j}{q_{i_n}-q_j}=0,
  \end{split}
\end{equation}
as the sum over $j\notin(i_1,\ldots,i_{n-1},i_n)$ gives zero, while
$q_{n-1}$ times the sum over $j\in(i_1,\ldots,i_{n-1})$ can be
replaced by $-\frac{n-1}{2}\qd_{i_n}\qd_{i_{n-1}}$. The searched
transformation $\qv\to\xv(q_1,\ldots,q_n)$ (non-singular, as long as
the $q_i$ are all different) is therefore provided by
\begin{align}
  &x_n[\qv\,] := \frac{1}{n!}\sumt q_{i_1}\cdots q_{i_n},\label{eq:xnsumt}\\
  &\det\para{\frac{\d x_n}{\d q_i}} = \prod_{i<j}(q_i-q_j),\\
  &\xdd_n[\qv(t)] = \dot{b}_n = 0.
\end{align}
The natural induced metric (giving \eqref{eq:Christoffel}) is 
\begin{equation}
  g_{ij} = \d_i\xv\cdot\d_j\xv = \para{J^TJ}_{ij},
\end{equation}
with
\begin{equation}
  J^n_j = \frac{\d x^n}{\d q^j}=\frac{1}{(n-1)!}\overset{\sim}{\sum_{\neq j}}
  q^{j_1}\cdots q^{j_{n-1}},
\end{equation}
(now writing the $q$ coordinates with upper indices) and the
goldfish-equations can therefore be described by the (geodesic flow)
Hamiltonian
\begin{align}
  &H[\qv,\piv]:=\half\pi_i\, g^{ij}[\qv\,]\pi_j,\\
  &g^{ij}=\para{J^{-1}(J^T)^{-1}}^{ij}=\frac{1+q^iq^j+\cdots+(q^iq^j)^{N-1}}{\prod_k'(q^i-q^k)\prod_l'(q^j-q^l)},\\
  &\qd^i = g^{ik}[\qv\,]\pi_k,\quad
  \pid_k = -\half\pi_i\frac{\d g^{ij}}{\d q^k}\pi_j
\end{align}
giving \eqref{eq:geodesic}/\eqref{eq:Christoffel}.

To the best of our knowledge, this geodesic interpretation (and
natural quadratic Hamiltonian structure) of the
Ruijsenaars-Schneider model has not been observed before.

The inverse of the Jacobian is
\begin{equation}
  \para{J^{-1}}^i_m=(-)^{m-1}\frac{(q^i)^{N-m}}{\prod_{j\neq i}(q^i-q^j)},
\end{equation}
and the conserved quantities
\begin{equation}
  B_n[\qv,\piv]:=\frac{1}{(n-1)!}\sumt q^{i_1}\cdots q^{i_{n-1}}g^{i_nj}[\qv\,]\pi_j
\end{equation}
(all linear in the momenta!) Poisson-commute, as the transformation
$(\xv,\pv)\leftrightarrow(\qv,\piv)$, with $p_n=\xd_n=B_n$ trivially gives $\pb{p_m,p_n}=0$.

Hamiltonian relations between zeroes of polynomials and their
coefficients have been considered before (cp.
\cite{Calogero:1997,Calogero:2001}), but -- to our suprise --
apparently \emph{not} for the original goldfish equation. We also
became aware of chapter 27 in \cite{Suris}\footnote{thanks to E.
  Langmann!}, where (just as in \cite{Calogero:2001}) the
time-independence of the quantities \eqref{eq:bn} is stated (and
proved), but with the standard exponential (cp. \eqref{eq:exp})
Hamiltonian structure, and not the geodesic structure (with respect to
which the conserved quantities Poisson-commute). Note that, rewriting
\cite{Suris} the canonical Poisson-bracket,
$\pb{q_i,\tilde{p}_j}=\delta_{ij}$, in terms of the exponential
variables suggested by the standard RS Hamiltonian
\begin{equation}\label{eq:exp}
  \sum_{i=1}^N e^{\tilde{p}_i}\prod_{j=1}^N(q_i-q_j)^{-1}=:
  \sum\pit_i = P,
\end{equation}
one has
\begin{equation}\label{eq:qpitbrackets}
  \pb{q_i,q_j}=0,\quad \pb{q_i,\pit_j}=\delta_{ij}\pit_i,\quad
  \pb{\pit_i,\pit_j}=\frac{\pit_i\pit_j}{q_i-q_j}(1-\delta_{ij})
\end{equation}
which (interpreting them as Dirac-brackets, in the context of our
Hamiltonian reduction to $H=\half\para{\sum p_i}^2$, with $p_i$
playing the role of $\pit_i$) tells one how to obtain the goldfish
equations directly from the reduced Hamiltonian \eqref{eq:HP}: using
\eqref{eq:qpitbrackets}, with $\pit_i$ replaced by $p_i$, and
\eqref{eq:exp} by $H=\half P^2$, gives \eqref{eq:goldfish}.

Finally, let us make some remarks about the hyperbolic case:
Starting, as was done for the Calogero-Moser case in \cite{Perelomov:1976}, with
\begin{align}\label{eq:XdXi}
  \frac{d}{dt}\para{\Xd\Xi+\Xi\Xd}=0,
\end{align}
$X$ a positive definite matrix -- which we take to be real, with
eigenvalues $e^{2a\lambda_j(t)}$ ($a$ real),
\begin{equation}\label{eq:RLambda}
  X(t) = R(t)e^{2a\Lambda(t)}\Ri(t),
\end{equation}
$R^T=\Ri$, one obtains (with, as before, $M:=\Ri\Rd$)
\begin{align}
  &L_{ij}=\delta_{ij}\ld_i-\frac{\sinh\para{2a(\lambda_i-\lambda_j)}}{2a}M_{ij},\\
  &\Ld = [L,M]\label{eq:Lax}
\end{align}
from \eqref{eq:XdXi}, when defining $L$ via 
\begin{equation}
  R(t)L(t)\Ri(t) = \frac{1}{4a}\para{\Xd\Xi+\Xi\Xd}.
\end{equation}
The solution of \eqref{eq:XdXi}, on the other hand, can be written as 
\begin{equation}
  X(t) = e^{a\Lambda_0}e^{2tV_0}e^{a\Lambda_0}
\end{equation}
(with $\Lambda_0=\Lambda(0)$ when choosing $R(0)=\mid$). The crucial
point is that \eqref{eq:Lax} consistently reduces to
\begin{equation}\label{eq:ldd}
  \ldd_i = 2\sump{j(\neq i)}\frac{2a\ld_i\ld_j}{\sinh(2a(\lambda_i-\lambda_j))}
\end{equation}
when setting 
\begin{equation}
  M_{ij} = -\frac{2a}{\sinh(2a(\lambda_i-\lambda_j))}\sqrt{\ld_i\ld_j}
\end{equation}
-- corresponding to the initial conditions
\begin{equation}\label{eq:V0ij}
  \para{V_0}_{ij} = a\frac{\sqrt{c_ic_j}}{\cosh a(a_i-a_j)},
\end{equation}
where $a_i=\lambda_i(0)$, $c_i=\ld_i(0)$ (geometrically speaking, the
matrix geodesics \eqref{eq:RLambda} project down to geodesics in the
space of positive diagonal matrices, if the initial conditions are
chosen ``perpendicular to the action of the rotation-group''). Except
for their intrinsically geodesic interpretation (made explicit below)
the equations \eqref{eq:ldd}, and variants thereof, are well-studied
(see e.g. \cite{Calogero:2001,Braden:1997}). In particular one can
show, that the solutions of
\begin{equation}
  \qdd_i = 2\sump{j}\qd_i\qd_j\coth(q_i-q_j)
\end{equation}
are the $N$ roots of 
\begin{equation}\label{eq:fq}
  f(q):=\sum_{i=1}^N\frac{\frac{c_i}{P}\tanh(Pt)}{\tanh(q-a_i)}-1=0,
\end{equation}
resp. (see e.g. \cite{Suris}) the eigenvalues of
$Z(t)=e^{2\Lambda_0}e^{2tL_0}$, with $P=\sum\qd_i=\sum c_i$,
$\para{L_0}_{ij}=c_j$ (avoiding the somewhat artificial positivity
restrictions in \eqref{eq:V0ij}). Rewriting \eqref{eq:fq} as
\begin{equation}
  \frac{d}{dt}\para{e^{-2tP}\frac{d}{dt}\det\para{Z(t)-e^{2q}\mid}}=0,
\end{equation}
the crucial step then (\emph{not} taken in \cite{Suris}) is to note that the symmetric functions of the eigenvalues $e^{2q_i(t)}$,
\begin{equation}
  s_n(t) = \frac{1}{n!}\sumt e^{2(q_{i_1}+\cdots+q_{i_n})}\qquad n=1,\ldots,N,
\end{equation}
in analogy with \eqref{eq:xnsumt}, evolve in time (almost uncoupled as
$2P=\Sd_N/s_N=\text{const.}$) as 
\begin{equation}
  \Sdd_n-2P\Sd_n=0\qquad n=1,\ldots,N.
\end{equation}

\section*{Acknowledgement}

\noindent We would like to thank the Swedish Research Council, the
Brainpool program of the Korea Research Foundation and the Korean
Federation of Science and Technology Societies, R14-2003-012-01002-0,
and the Marie Curie Training Network ENIGMA, for support.

\newpage

%%
%% Bibliography
%%

\bibliographystyle{plain}

\end{document}